\documentclass[a4paper,11pt]{article}
\usepackage{multirow}
\usepackage{amsfonts}
\usepackage{adjustbox}
\usepackage{amsmath}
\usepackage{graphicx}
\usepackage{pict2e}
\usepackage{epsfig}
\usepackage[
    font      = footnotesize,
    labelfont = bf
]{caption}
\usepackage[
    font      = footnotesize,
    labelfont = bf
]{subcaption}

\usepackage[colorlinks=true]{hyperref}

\begin{document}
\newcommand{\ds}{\displaystyle}
\newcommand{\be}{\begin{equation}}
\newcommand{\ee}{\end{equation}}
\newcommand{\ba}{\begin{array}}
\newcommand{\ea}{\end{array}}
\newcommand{\bea}{\begin{eqnarray}}
\newcommand{\eea}{\end{eqnarray}}

\def\R{{\mathbb R}}
\def\C{{\mathbb C}}
\newcommand{\bi}{\begin{itemize}}
\newcommand{\ei}{\end{itemize}}
\newcommand{\sn}{\mbox{sn}}
\newcommand{\cn}{\mbox{cn}}
\newcommand{\dn}{\mbox{dn}}
\newcommand{\x}{{\ensuremath{\times}}}
\newcommand{\bb}[1]{\makebox[16pt]{{\bf#1}}}
\newtheorem{theorem}{Theorem}
\newtheorem{definition}{Definition}
\newtheorem{lemma}{Lemma}
\newtheorem{comment}{Comment}
\newtheorem{corollary}{Corollary}
\newtheorem{example}{Example}
\newtheorem{examples}{Examples}
 \newtheorem{conjecture}{Conjecture}
\title{Conformal Laplace superintegrable systems in 2D: polynomial invariant subspaces}
\author{{\bf M.A. Escobar-Ruiz}\\
{\sl Instituto de Ciencias Nucleares, UNAM}\\
{ \sl Apartado Postal 70-543, 04510 Mexico D.F. MEXICO}\\
{\sl and School of Mathematics, University of Minnesota}\\
{\sl mauricio.escobar@nucleares.unam.mx}\\
{\bf  and  Willard Miller, Jr.}\\
{\sl School of Mathematics, University of Minnesota,}\\
{\sl Minneapolis, Minnesota,
55455, U.S.A.}\\
{\sl miller@ima.umn.edu}}
\date{\today}
\date{\today}
\maketitle
%\tableofcontents

\begin{abstract}

2nd-order conformal superintegrable systems in $n$ dimensions are Laplace equations on a manifold with an added scalar potential and $2n-1$ independent 2nd order conformal
symmetry operators. They encode all the information
about Helmholtz (eigenvalue)  superintegrable systems in an efficient manner: there is a 1-1 correspondence between Laplace superintegrable systems
and St\"ackel equivalence classes of Helmholtz superintegrable systems. In this paper we focus on superintegrable systems in two dimensions, $n=2$,
where there are 44 Helmholtz systems, corresponding to 12 Laplace
systems. For each Laplace equation we determine the possible  $2-$variate polynomial subspaces that are invariant under the action of the
Laplace operator, thus leading to families of polynomial eigenfunctions.
 We also study the behavior of the polynomial
invariant subspaces under a St\"ackel transform.
The principal new results are the details of the polynomial variables and the conditions on parameters of the potential corresponding to
polynomial solutions. The hidden $gl_3$-algebraic structure is exhibited for the exact and quasi-exact systems.
For physically meaningful solutions, the orthogonality properties and normalizability of the polynomials are presented as well.
Finally, for all Helmholtz superintegrable solvable systems we give a unified construction of 1D and 2D quasi-exactly solvable potentials possessing polynomial solutions, and a construction of new 2D PT-symmetric potentials is established.
\end{abstract}

\section{Introduction}

This paper is part of a series  \cite{BocherCon1, BocherCon2} whose purpose is to systematize and unify the  study of 2nd order
Helmholtz (i.e. Schr\"odinger eigenvalue)  superintegrable systems in 2D and 3D
by transforming these systems to conformally superintegrable Laplace equations with a scalar potential
and applying ideas originally due to B\^ocher, \cite{Bocher}.  It is well
known \cite{KKMP2011,KKM20042, KKM20061,MPW2013,Kress2001, CK2014,Kress2007}
that any second order
conformal Laplace superintegrable system admitting a non-constant
potential $V({\bf x})$ can be St\"ackel transformed to a Helmholtz
superintegrable system, and this  operation is invertible. Moreover,
each  family of St\"ackel-equivalent Helmholtz superintegrable
systems on a variety of manifolds corresponds to a {\it single}
conformally superintegrable system on flat space. The different
Helmholtz systems in an equivalence class share some important
properties. For example, their structure algebras are isomorphic,
modulo a permutation of parameters. Furthermore, by taking a gauge transformation, most of these
Laplace equations can be transformed into an eigenvalue problem with a Hamiltonian operator
that leaves invariant a polynomial vector space. This polynomial vector
space is the same
for  all Hamiltonians in a St\"ackel equivalence class and it can most
conveniently be studied via Laplace equations.
Except for a few exceptional special cases, these Helmholtz
superintegrable systems are multiseparable, with the various separable
solutions characterized as eigenfunctions of a 2nd order symmetry
operator for the 2D systems. Typically this operator also leaves
the polynomial vector space invariant, and the eigenfunctions are
polynomial special functions. The Hamiltonian and its 2nd order
symmetries are formally self-adjoint so this construction leads to
families of orthogonal polynomials. The usual hypergeometric
orthogonal polynomials such as  Jacobi, Laguerre and Hermite arise in
this way, but also non-hypergeometric polynomials such as Heun and
spheroidal appear. This is one important way that special functions and
orthogonal polynomials are related to superintegrable systems.
These special functions all satisfy differential equations. A second
important way special functions are related to such systems is due
to multiseparability. The separated eigenfunctions
characterized by one symmetry operator can be expanded in the eigenbasis
of another symmetry operator. The expansion coefficients are
themselves special functions with orthogonality properties, and they may
satisfy difference as well as differential equations.
For example, Wilson and Racah polynomials arise in this way, e.g.
\cite{KMP2007a,LM2014}. The contraction scheme relating superintegrable systems leads
to limit relations for orthogonal polynomials and special functions,
including the Askey Scheme for hypergeometric orthogonal polynomials,
\cite{KMP2014,KM2014}.

Here we consider all 2D 2nd order superintegrable systems   and
determine when systems are exactly solvable or quasi-exactly solvable,
i.e., we
find the gauge factor and the variables for which the gauge transformed
operator possesses polynomial invariant subspaces. We also pay attention
to the issue of when the 1D separation equations which arise from these
systems are QES and the polynomial special functions that occur.
Finally, since all of these
systems have classical mechanical counterparts we discuss the relations
between the trajectories and their time dependence for classical Helmholtz
superintegrable systems that belong to the same equivalence class.

\subsection{Conformally superintegrable Laplace systems}

We consider Laplace systems of the form
  \be\label{Laplace} H\,\Psi({\bf x}) \ \equiv \ (\,\Delta_2+V({\bf
x})\,)\,\Psi({\bf x}) \ =\ 0\ .\ee
Here $\Delta_2 $ is the Laplace-Beltrami operator on a real or complex
$2D$ Riemannian or pseudo-Riemannian manifold and $V$ is
a  non-zero scalar potential. All variables can be complex, except when we impose constraints such as square integrability.
A conformal symmetry of  equation (\ref{Laplace}) is a partial
differential operator  $L$ in the variables ${\bf x}=(x_1,\,x_2)$ such
that $[ L, H]\equiv LH-HL=R_{ L} H$ for some differential operator $R_{L}$.
A conformal symmetry maps any solution $\Psi$ of (\ref{Laplace}) to
another solution. Two conformal symmetries ${ L}, { L}'$ are
identified if $L=L'+R\,H$ for some differential operator $R$, since they
agree on the solution space of (\ref{Laplace}). (For brevity we will say
that $L=L', \mod (H)$
and that $L$ is a symmetry if $[L,H]=0,\mod(H)$.)
The system is {\it conformally superintegrable} if there exist three
algebraically independent conformal symmetries,
${ L}_1,\,{L}_{2},\, L_3$ with ${L}_3={ H}$. It is second order
conformally superintegrable if each
$L_2$ can be chosen to be a 2nd order differential operator, and $L_1$
of at most 2nd order. (If the system admits symmetries such that
$L_1,L_2$ can be chosen as 1st order, we
say it is 1st order conformally superintegrable). Recall that a
Helmholtz eigenvalue system
${\hat H}\Psi=E\Psi$ with ${\hat H}=\Delta  +{\hat V}$ is {\it
superintegrable } if there exist three algebraically independent true
symmetries,
${\hat  L}_1,\,{\hat L}_{2},\,{\hat L}_3$ with ${\hat L}_3={\hat H}$,
i.e., $[{\hat L_j},{\hat H}]=0$.

  Every 2D Riemannian manifold is conformally flat, so we can always find
a Cartesian-like coordinate system with coordinates $(x,\,y)\equiv
(x_1,\,x_2)$ such that  a Helmholtz  system takes the form
  \[{\hat
H}\,\Psi=\frac{1}{\lambda(x,y)}(\partial^2_{x}+\partial^2_{y}+{\hat
V}(x,y))\,\Psi\ =\ E\,\Psi.\]
  We can rewrite this as a flat space Laplace system $H\Psi=0$ where
$H=\partial_x^2+\partial_y^2 +V$ with $V={\hat V}-E\lambda$. It is easy
to show
  that this Laplace system is conformally superintegrable if and only if
the Helmholtz system is superintegrable. Thus any Helmholtz superintegrable
  system on any manifold corresponds to a flat space Laplace conformally
superintegrable system. Moreover, given a flat space
  Laplace superintegrable system with metric  $ds^2=dx^2+dy^2$, measure
$dx\, dy$ and
  potential of  the form  $V=V_0+\alpha\, U$ for functions $V_0,U$ and
parameter $\alpha$,  then its {\it Conformal St\"ackel Transform }
   \[ CST:\quad {\tilde H}\,\Psi\ =\
U^{-1}\left(\partial^2_{x}+\partial^2_{y}+V_0\right)\Psi\ =-\alpha\Psi\]
  is a Helmholtz superintegrable system with metric $ds^2=U(dx^2+dy^2)$,
measure $ U\ dx\ dy$.
There is an analogous definition of St\"ackel transforms that take one
Helmholtz superintegrable system to another, \cite{KKM20042}. We see from this
that each equivalence class
of Helmholtz superintegrable systems corresponds to a single conformally
superintegrable flat space Laplace system.

In papers \cite{BocherCon1,BocherCon2}  it is shown that the 44 families of Helmholtz
superintegrable systems correspond to exactly 14 Laplace systems.
They are of two types: the 8
systems with non-degenerate potentials (4 parameter, 3 parameter for the
Helmholtz systems), Table \ref{Tab1}, and  the 6 systems with degenerate
potentials (2 parameter,
1 parameter for the Helmholtz systems), Table \ref{Tab2}. There are no
other possibilities. The degenerate potentials can all be obtained as
parameter restrictions of nondegenerate ones, but they have additional
symmetry not inherited from the restriction. All of the nondegenerate
Laplace systems can be obtained as
B\^ocher contractions of system $[1,1,1,1]$, \cite{BocherCon2}. All of the degenerate Laplace systems
can be obtained as B\^ocher contractions of system $A$.

{
\begin{table}[t!]
\begin{center}
\resizebox{\textwidth}{!}{%
\begin{tabular}{|c||c| }
\hline
   &   \\
\multirow{2}{*}{System}  &    Non-degenerate potentials   \\ [1.5ex]
     &$V(x,\,y)$\\  [2.5ex]
\hline
\hline
  &   \\
$ [1,1,1,1] $ & $
\frac{a_1}{x^2}+\frac{a_2}{y^2}+\frac{4\,a_3}{(x^2+y^2-1)^2}-\frac{4\,a_4}{(x^2+y^2+1)^2}

$
\\ [1.8ex]
$ [2,1,1]  $ & $ \frac{a_1}{x^2} + \frac{a_2}{y^2} - a_3\,(x^2+y^2) + a_4  $
\\  [1.8ex]
$ [2,2]    $ & $
\frac{a_1}{(x+i\,y)^2}+\frac{a_2\,(x-i\,y)}{(x+i\,y)^3}+a_3-a_4\,(x^2+y^2) $
\\  [1.8ex]
$ [3,1]    $ & $ a_1-a_2\,x+a_3\,(4\,x^2+y^2)+\frac{a_4}{y^2} $
\\  [1.8ex]
$ [4]      $ & $ a_1-a_2\,(x+i\,y)
+a_3\,(3(x+i\,y)^2+2(x-i\,y))-a_4\,(4(x^2+y^2)+2(x+i\,y)^3) $
\\  [1.8ex]
$ [0]      $ & $ a_1-(a_2\,x+a_3\,y)+a_4\,(x^2+y^2)   $
\\  [1.8ex]
$ (1)      $ & $ \frac{a_1}{(x+i\,y)^2}+a_2
-\frac{a_3}{(x+i\,y)^3}+\frac{a_4}{(x+i\,y)^4} $
\\  [1.8ex]
$ (2)      $ & $ a_1+a_2(x+i\,y)+a_3(x+i\,y)^2+a_4(x+i\,y)^3 $
\\  [2.5ex]
\hline
\end{tabular}}
\caption{Four-parameter potentials. Each
of the Helmholtz nondegenerate superintegrable
(i.e. 3-parameter) eigenvalue systems is St\"ackel equivalent
to exactly one of these systems. Thus, with
one caveat mentioned in \cite{BocherCon1}, there are exactly 8
equivalence classes of
Helmholtz systems.}
\label{Tab1}
\end{center}
\end{table}
}
\begin{table}[h!]
\begin{center}
\small
\setlength{\tabcolsep}{2.0pt}
\begin{tabular}{|c||c| }
\hline
   &   \\
\multirow{2}{*}{System}  &    Degenerate potentials   \\ [1.0ex]
     &  $V(x,\,y)$\\  [2.0ex]
\hline
\hline
  &   \\
$ A $ & $ \frac{4\,a_3}{(x^2+y^2-1)^2}-\frac{4\,a_4}{(x^2+y^2+1)^2} $
\\ [1.5ex]
$ B  $ & $ \frac{a_1}{x^2}+a_4 $
\\  [1.5ex]
$ C    $ & $ a_3-a_4\,(x^2+y^2) $
\\  [1.5ex]
$ D    $ & $ a_1-a_2\,x $
\\  [1.5ex]
$ E      $ & $ \frac{a_1}{(x+i\,y)^2}+a_3  $
\\  [1.5ex]
$ F     $ & $ a_1-a_2\,(x+i\,y)  $
\\  [2.0ex]
\hline
\end{tabular}
\caption{Two-parameter degenerate potentials.}
\label{Tab2}
\end{center}
\end{table}

\section{Relations between conformal St\"ackel transforms and polynomial solutions of Laplace systems}

For an eigenvalue equation ${\cal H} \,\psi({\bf x}) \ = \ a\,\psi({\bf x})$ with spectral parameter $a$:

\begin{itemize}
  \item The operator $\cal H$ is said to be \textbf{exactly-solvable}, \textbf{(ES)} if there exists an infinite flag
  of subspaces ${\cal P}_{N}$, $N=1,2,3,...,$ such that $n_N=\text{dim}{\cal P}_{N}\rightarrow \infty$ as $N\rightarrow \infty$ and
  ${\cal H}\,{\cal P}_{N}\subseteq {\cal P}_{N}\subseteq {\cal P}_{N+1}$ for any $N$. In this case, for each
  subspace ${\cal P}_{N}$ the $n_N$ eigenvalues and eigenfunctions of $\cal H$ can be obtained by pure algebraic means.

  \item The operator $\cal H$ is called \textbf{quasi-exactly solvable}, \textbf{(QES)} if there exist a single subspace
${\cal P}_{k}$ of dimension $n_k>0$  such that ${\cal H}\,{\cal P}_{k}\subseteq {\cal P}_{k}$. In this case, again we can find $n_k$
eigenvalues and eigenfunctions of $\cal H$ by algebraic means, but we have no information about the remaining eigenvalues and eigenfunctions.
\end{itemize}
See \cite{TTW:2001, TurbinerQES2,Ushveridze, ExactAndQES, Turbiner:1988}.

We will consider the 14 Laplace 2D  systems $  H\,\Psi\ =\ (\,\partial^2_{x}+\partial^2_{y}+V\,)\,\Psi \ = \ 0$,
8 with non-degenerate potential $V\,=\,V({\bf x};\,a_1,\,a_2,\,a_3,\,a_4)$,
which depend linearly on  four parameters, and 6 systems with  degenerate potential, depending linearly on two parameters.
We will show that, after an appropriate  gauge transformation, most of these Laplace systems lead to an
eigenvalue problem which is exactly solvable or quasi-exactly solvable, as are all Helmholtz systems obtained from them by
conformal St\"ackel transforms.

In particular, we will show that most of the Laplace systems mentioned above possess a hidden $gl_3$ algebra realized by the generators \cite{Turbiner:1988}
\begin{equation}
\begin{aligned}
& {\cal J}_i^- =\partial_{w_i}   \ , \qquad i=1,2\, ,
\\ &  {\cal J}_{ij}^0  =  w_i\,\partial_{w_j} \ , \qquad i,j=1,2 \, ,
\\ &  {\cal J}^0(N) = w_1\,\partial_{w_1}+w_2\,\partial_{w_2} -N  \ ,
\\ &  {\cal J}_i^+(N)=w_i\,{\cal J}^0(N)= w_i\,(w_1\,\partial_{w_1}+w_2\,\partial_{w_2} -N)  \ .
\label{generators}
\end{aligned}
\end{equation}
The parameter $N$ in (\ref{generators}) can be any real number. However, if $N$ is a non-negative integer, the representation (\ref{generators}) of the $gl_3$ algebra becomes the finite-dimensional representation acting on the space of polynomials
\bea\label{InvSub}  {\cal P}_{N}^{(2)} \ =\ \langle  w_1^{p_1}\,w_2^{p_2}\mid 0\leq p_1+p_2\leq  N \rangle   \ , \eea
which form  the flag
${\cal P}_0^{(2)}\subset {\cal P}_1^{(2)} \subset {\cal P}_2^{(2)}...\subset {\cal P}_N^{(2)} \subset ...\,{\cal P}$.
A quadratic function in the generators (\ref{generators}) maps the polynomial space (\ref{InvSub}) into itself.

For exactly and quasi-exactly solvable systems we will determine the following three elements
\begin{enumerate}
  \item The polynomial variables $w_{j}=w_{j}({\bf x})$, $j=1,2$. They are the variables in which the  (\ref{Laplace})
   Laplace system admits invariant domains where $\Psi$ takes the form of a polynomial $P$ (of total order $N$) times a common (gauge)
   factor $\Psi_0$,
   \[ \Psi \ =\ \Psi_0(w_1,\,w_2)\,P_N(w_1,\,w_2)   \ . \]
   For the polynomial systems we will find that, except in $[1,1,1,1]$ and $A$ listed in Tables \ref{Tab1} and \ref{Tab2} respectively, factorization of variables occurs: $\Psi_0(w_1,\,w_2)=\psi_1(w_1)\,\psi_2(w_2)$ and $P_N(w_1,\,w_2)=p_1(w_1)\,p_2(w_2)$.
  \item The gauge factor $\Psi_0(w_1,\,w_2)$. It plays the role of a generalized ``ground state function". This non-polynomial function $\Psi_0(w_1,\,w_2)$ describes the asymptotic behavior of $\Psi$ in (\ref{Laplace}), i.e., it determines whether the solution is square integrable or not.
  \item The general constraint. The operator $H$ in the original Laplace equation $H\,\Psi=0$ is gauge transformed to
  \be \label{Htrans} h \ \equiv \  \Psi_0^{-1}\,H\,\Psi_0   \ .  \ee
  In general, the condition $H\,\Psi=0$ leads to an eigenvalue problem $h'\,P \ = \ E\,P$,
  where $E=E(a_1,a_2,a_3,a_4)$ and $h'=h-E$. Unlike $h$, the operator $h'$ has no constant term.
  Moreover, the operator $h$ (\ref{Htrans}) maps polynomials into polynomials without increasing the order. Therefore,
  it can be written in terms of generators (\ref{generators}) where $N$ in (\ref{generators}) is determined by the parameters of the
  potential $V$: $N\ =\  N(a_1,\,a_2,\,a_3,\,a_4)$.
  Then, for a non-negative integer $N$ the operator $h$ (\ref{Htrans}) possesses a polynomial invariant subspace ${\cal P}_N$, i.e.,
 in variables $w_1, w_2$.
\end{enumerate}

\subsection{Summary of relations between conformal St\"ackel transforms and  polynomial solutions of Laplace systems}
\begin{itemize}
 \item Conformal St\"ackel Transform (CST): Assume
 \[ H\Psi=(\partial^2_{x}+\partial^2_{y}+V(x,y))\Psi=0;\quad V=V_0+\alpha U,\]
 \[ {\rm metric}:\ ds^2=dx^2+dy^2,\quad {\rm measure}:\ dx\ dy,\]
 \[ CST:\quad {\tilde H}\Psi=\left(U^{-1}(\partial^2_{x}+\partial^2_{y})+U^{-1}V_0+\alpha\right)\Psi=0,\]
 \[ {\rm metric}:\ ds^2=U(dx^2+dy^2),\quad {\rm measure}:\ U\ dx\ dy.\]
 \item Polynomial solutions: Assume $H\Psi =0$ with $\Psi = \Psi_0(x,y)P(w_1,w_2)$ where $\Psi_0$ is the ``ground state wave function'' and
 $P$ is a polynomial in suitable independent variables $w_j(x,y)$, $j=1,2$.
 \[ H'P=\left(\Psi_0^{-1}(\partial^2_{x}+\partial^2_{y})\Psi_0+V_0+\alpha U\right ))P=0,\]
 \[ \Psi_0^{-1}(\partial^2_{x}+\partial^2_{y})\Psi_0 =\partial^2_{x}+\partial^2_{y}
  +2(\,(\partial_x\Psi_0)\partial_x+(\partial_y\Psi_0)\partial_y\,) +(\partial^2_{x}\Psi_0)+(\partial^2_{y}\Psi_0) \ .\]
\[  {\rm measure}:\ {|\Psi_0|}^2\ dx\ dy.\]
  \[ CST:\quad {\tilde H}'P=\left((\Psi_0\,U)^{-1}(\partial^2_{x}+\partial^2_{y})\Psi_0+U^{-1}V_0+\alpha\right)P=0,\]
  \[ {\rm metric}:\ ds^2=U(dx^2+dy^2),\quad {\rm measure}:\ {|\Psi_0|}^2\ U\ dx\ dy.\]
\end{itemize}

\section{Non-degenerate potentials}

In this section we treat the  8 generic Laplace superintegrable systems with non-degenerate potential, listed in Table \ref{Tab1}. Each
of the $44$ known Helmholtz non-degenerate superintegrable
(i.e. 3-parameter) eigenvalue systems is St\"ackel equivalent to exactly one of these generic systems, \cite{BocherCon2}.

\begin{enumerate}
\item System $[1111]$:

This system is exactly solvable and is $R$-separable in 2 sets of coordinates
\bea\label{spherical} &(a)&\ {\rm Spherical:}\
x =\frac{ \sin \theta \cos \phi}{1+\cos\theta},\ y = \frac{\sin \theta \sin \phi}{1+\cos\theta},  \\
 &(b)&\ {\rm  Elliptic:}\quad\  x^ 2 = \frac{(c\,u - 1)(c\,v - 1)}{(1 - c)(1+\sqrt{c\,u\,v})^2},\ y^2=\frac{c(u-1)(v-1)}{(c-1)(1+\sqrt{c\,u\,v})^2}, \nonumber \\
& &\qquad c \ {\rm is\ a\ parameter}\ \ne 0,1. \nonumber  \eea
The polynomial variables are
\[ w_1\ =\ \frac{4\,x^2}{(x^2+y^2+1)^2}\ ,\qquad w_2\ =\  \frac{4\,y^2}{(x^2+y^2+1)^2}.\]
Writing the function $\Psi(w_1,w_2)=\Psi_0(w_1,w_2)\,P(w_1,w_2)$ in (\ref{Laplace}) with the gauge factor
$\Psi_0=w_1^{k_1}w_2^{k_2}(1-w_1-w_2)^{k_3}$,
and
\[ a_1=-2\,k_1\,(2\,k_1-1),\ a_2=-2\,k_2\,(2\,k_2-1),\ a_3=-2\,k_3\,(2\,k_3-1), \]
we arrive at the  equation for the polynomial function $P(w_1,w_2)$: {\small
\bea\label{P1111} h^{(ES)}\,P\equiv  \bigg[ 2\,w_1\,(1-w_1)\,\partial_{w_1}^2+2\,w_2\,(1-w_2)\partial_{w_2}^2-4\,w_1\,w_2\,\partial_{w_1,w_2}^2 \eea
\[ +\left(1+4k_1-w_1(3+4\,K)\right)\partial_{w_1}+\left(1+4k_2-w_2(3+4\,K)\right)\partial_{w_2}+E_0\bigg]\,P \, =\, 0\ , \] }
where $K\equiv k_1+k_2+k_3$ and $E_0=-\frac{1}{2}[ a_4  +   2\,K(1+2\,K)  ] $. Variables $w_1,\,w_2$ in (\ref{P1111}) are not separated,
but the operator $h^{(ES)}$ in (\ref{P1111}) acting on $P$ maps polynomials, in these variables, into polynomials without increasing the order and is formally self-adjoint with respect to the inner product
 {\small \[ \langle P_1 , P_2 \rangle \equiv \int \int \, dw_1\,dw_2\, P_1(w_1,w_2)\,
 \overline{P_2}(w_1,w_2)\,|\Psi_0|^2\,w_1^{-\frac{1}{2}}\,w_2^{-\frac{1}{2}}\,{(1-w_1-w_2)}^{-\frac{1}{2}} \ .  \]}
Boundaries of the configuration space (domain), $w_1\geq0,\,w_2\geq0$ and $w_1+w_2\leq1$, are determined by zeros of $\Psi_0$.
Square-integrability demands $k_1,k_2,k_3\geq {1}/{4}$.

Equation (\ref{P1111}) can be rewritten in terms of $gl_3$ generators (\ref{generators}): {\small
\[h^{(ES)}\,P\ =\ \left[ 2{\cal J}^0_{11}({\cal J}^-_{1}-{\cal J}^0_{11})  + (1+4\,k_1){\cal J}^-_{1} -(1+4\,K){\cal J}^0_{11} \right.\ \]
\[ \left. + 2\,{\cal J}^0_{22}({\cal J}^-_{2}-{\cal J}^0_{22})  + (1+4\,k_2){\cal J}^-_{2}   -(1+4\,K){\cal J}^0_{22} -4{\cal J}^0_{12}
+E_0\right]P=0 \ .\]}
Thus the operator $h^{(ES)}$ has a hidden $gl_3$ algebra. Infinitely many finite-dimensional invariant subspaces (\ref{InvSub}) leads to the constraint
\begin{equation}
2\,(N+K)(2\,N+2\,K+1) + a_4\ = \ 0   \ ,
\label{polyconstraint1}
\end{equation}

By adding the trivial raising operator
\[ {\cal J}^+(N)\ \equiv \ {\cal J}_1^+(N) +{\cal J}_2^+(N) \ = \ (w_1+w_2)(w_1\,\partial_{w_1}+w_2\,\partial_{w_2}-N)   \ ,  \]
the exactly-solvable operator $h^{(ES)}$ that annihilates $P$ can be easily generalized to a quasi-exactly-solvable one
$h^{(QES)}=h^{(ES)} + \alpha\,{\cal J}^+(N)$,
with real parameter $\alpha$. Now this operator has a single invariant subspace in $2-$variate polynomials. Moreover, with a term
$I = {\cal J}^-_{1}-{\cal J}^-_{2}$,
we can form the operator $h^{(PT)}$
\[ h^{(PT)}=h^{(ES)} + \alpha\,{\cal J}^+(N) + \beta\,I \ , \]
where $\beta\neq 0$ is a parameter. The operator $h^{(PT)}$ is also quasi-exactly-solvable. However, unlike $h^{(ES)}$ (\ref{P1111}), $h^{(PT)}$ is not formally
self-adjoint due to the boundary terms. For $\beta$ a pure complex number $h^{(PT)}$ is invariant under the operation of complex
conjugation followed by the transposition $w_1\leftrightarrow w_2$. Thus the system is PT-symmetric, with real energy eigenvalues, \cite{Mos}.

The basis of conformal symmetries is given by the set $\{ L_1,\,L_2,\,h^{(ES)}\}$ where {\small
\bea L_1\ &=\ &w_1\,w_2(\partial_{w_1}^2-2\partial_{w_1\,w_2}+\partial_{w_2}^2)+
\left(\frac{w_1}{2}(1+4k_2)-\frac{w_2}{2}(1+4k_1)\right)(\partial_{w_2}-\partial_{w_1}),\nonumber\\
 L_2\ &=\ &-w_1\,(w_1+w_2-1)\partial_{w_1}^2+\nonumber\\
 &&\left(\frac12+2\,k_1-w_1(1+2\,k_1+2\,k_3)-\frac12w_2(1+4k_1)\right)\partial_{w_1}-2\,k_1.\nonumber\eea}
which also can be written in terms of $gl_3$ generators (\ref{generators})
\small{ \[   L_1 \ = \  {\cal J}^0_{12}\,{\cal J}^0_{22}+{\cal J}^0_{21}\,{\cal J}^0_{11}-2{\cal J}^0_{22}\,{\cal J}^0_{11}-{\cal J}^0_{12}-{\cal J}^0_{21}   \] \[       +  \frac{1+4\,k_2}{2}({\cal J}^0_{12}-{\cal J}^0_{11}) + \frac{1+4\,k_1}{2}({\cal J}^0_{21}-{\cal J}^0_{22}) \ ,  \nonumber \]}
\small{ \[   L_2 \ = \  -{\cal J}^0_{21}\,{\cal J}^0_{11}-{\cal J}^0_{11}\,{\cal J}^0_{11}+{\cal J}^0_{11}\,{\cal J}^-_{1}
+{\cal J}^0_{21}+{\cal J}^0_{11}  +  (\frac{1}{2}+2\,k_1)({\cal J}^-_{1}\]
\[-{\cal J}^0_{21})-(1+2k_1+2k_3){\cal J}^0_{11}-2\,k_1 \ .   \nonumber \]}
The solutions of (\ref{P1111}) separable in spherical coordinates are products of Jacobi polynomials and are the eigenfunctions of $L_1$ while those separable in elliptic coordinates are eigenfunctions of $L_1+(1-r)L_2$ and can be factorized as the the product of two Heun polynomials.

Even though this is an exactly solvable system, the separation equations in elliptic coordinates are quasi-exactly-solvable. Indeed, setting
$P(u,v)\,=\,U(u)\,V(v)$ with the constraint (\ref{polyconstraint1}) we find the separation equation {\small
 \[4u(ru-1)(u-1)U''(u)+2\left((4rK+3r)u^2-(4rK+2r+2+4k_2+4k_3)u\right.\]
 \[\left. +4k_3+1\right)U'(u)
 +2rN(2N+4K+1)u U(u) =\Lambda U(u),\]}
 with an exactly similar equation for $V(v)$,
 where $\Lambda
 $ is the separation parameter. These equations are QES; they have $N+1$ eigenvalues $\Lambda_j$ on the space of polynomials of maximum order $N$ in $u$ and $v$,
 respectively.
For Helmholtz versions of these observations, see \cite{ExactAndQES}.

\item System $[211]$:

The system $[211]$ corresponds to the well-known Smorodinsky-Winternitz potential \cite{SWpotential}. It is $R$-separable in 3 sets of coordinates
 \bea\label{cartesian} &(a)&\ {\rm Cartesian:}\qquad x,\ y,\\
 &(b)&\ {\rm  Polar:}\quad\quad \ x=r\,\cos\theta,\ y=r\,\sin\theta,\nonumber\\
 &(c)&\ {\rm Elliptic:}\quad\  x=c\,\sqrt{(u-1)(v-1)},\ y= c\,\sqrt{-u\,v}\ ,\nonumber
 \eea
 where $c\neq 0$ is a parameter.
The polynomial variables are
$ w_1=x^2,\ w_2=y^2$,
and the gauge factor reads
$ \Psi_0=\exp\left[k_3(w_1+w_2)\right] w_1^{k_1}w_2^{k_2}$. With
\[ a_1=-2k_1(2k_1-1),\ a_2=-2k_2(2k_2-1),\ a_3=4k_3^2,\ \ ,  \]
and $\Psi(w_1,w_2)=\Psi_0(w_1,w_2)\,P(w_1,w_2)$ we obtain the equation
\[ h^{(ES)}\,P\equiv[ 4w_1\partial_{w_1}^2+4w_2\partial_{w_2}^2+
 2(4k_1+1+4k_3w_1)\partial_{w_1}\]
\bea\label{P211}+2(4k_2+1+4k_3w_2)\partial_{w_2}  +  4\,k_3\,(1+2\,k_1+2\,k_2)+a_4 ]\,P=0. \eea

Unlike the previous case (\ref{P1111}), in this problem separation in the variables $w_1,\,w_2$ occurs.
Factorization takes place and, essentially, we are left with two independent one-dimensional problems. The operator
$h^{(ES)}$ maps polynomials, in variables $w_1,\,w_2$, into polynomials without increasing the order and is formally
self-adjoint with respect to the inner product
\[ \langle P_1 , P_2 \rangle \equiv \int \int \, dw_1\,dw_2\, P_1(w_1,w_2)\,\overline{P_2}(w_1,w_2)\, |\Psi_0|^2\, w_1^{-\frac{1}{2}}\,w_2^{-\frac{1}{2}} \ .  \]
Boundaries of the configuration space (domain), $w_1\geq0,\,w_2\geq0$ are determined by zeros and singularities of $\Psi_0$. Square-integrability demands $k_1\geq \frac{1}{4},\,k_2\geq \frac{1}{4}$,$k_3< 0$.

The basis of symmetries is $\{ L_1,\,L_2,\,h^{(ES)}\}$ where {\small
 \bea  L_1 & =&  2w_1\partial_{w_1}^2+(4k_1+1+4k_3w_1)\partial_{w_1}+2k_3(1+2k_1),\nonumber\\
  L_2 &=& 4w_1w_2(\partial_{w_1}^2-2\partial_{w_1w_2}+\partial_{w_2}^2)+(-8k_2w_1+8k_1w_2-2w_1+2w_2)\partial_{w_1}, \nonumber\\
 &+&(8k_2w_1-8k_1w_2+2w_1-2w_2)\partial_{w_2}.\nonumber\eea }
The separation equations in Cartesian and polar coordinates are exactly solvable (products of associated
Laguerre polynomials in the Cartesian case, products of associated Laguerre and Jacobi polynomials in the polar case);
for elliptic coordinates they are QES (products of Heun polynomials). Indeed, setting
 $P(u,v)=U(u)V(v)$ with the constraint
\be\label{polyconstraint2}  4\,k_3\,(2\,N+2\,k_1+2\,k_2+1)+a_4=0.\ee
 we find the separation equation {\small
 \[-4u(u-1)U''(u)+2\left(4k_3c^2u^2-2(2k_1+2k_2+2k_3^2+1)u+4k_2+1\right)U'(u)\]
 \[-8\,k_3\,c^2\,N \,u\, U(u)\ =\ \Lambda \ U(u),\]}
 with an exactly similar equation for $V(v)$,
 where $\Lambda
 $ is the separation parameter. These equations have $N+1$ eigenvalues $\Lambda_j$ on the space of polynomials of
 maximum order $N$ in $u$ and $v$, respectively.

In terms of the generators ${\cal J}$`s (\ref{generators}) the equation (\ref{P211}) takes the form
\[h^{(ES)}\,P\equiv\left[ 4\,{\cal J}^0_{11}\,{\cal J}^-_{1} +8\,k_3\,({\cal J}^0_{11}+{\cal J}^0_{22})+2\,(1+4\,k_1){\cal J}^-_{1} \right. \]
\[\left. + 4\,{\cal J}^0_{22}\,{\cal J}^-_{2} +2(1+4\,k_2){\cal J}^-_{2} + E_0     \right]P=0 \  ,\]
If the constraint (\ref{polyconstraint2}) is satisfied then $h^{(ES)}$ admits a basis of polynomial solutions of
highest total order $N=0,1,2,...$ in variables $w_1,w_2$.
The dependence of the spectral parameter $-a_4$ is linear in  quantum number $N$.

Also, $h^{(ES)}$ becomes quasi-exactly-solvable by adding the raising operator ${\cal J}^+(N)$:
$h^{(QES)}=h^{(ES)} + \alpha\,{\cal J}^+(N)$,
with parameter $\alpha$. The term
$I = {\cal J}^-_{1}-{\cal J}^-_{2}$,
$ h^{(PT)}=h^{(ES)} + \alpha\,{\cal J}^+(N) + \beta\,I$,
with $\beta$ a parameter can be added as well and the system remains quasi-exactly-solvable.
However, for $\beta\neq0$, due to the boundary terms it is not formally self-adjoint anymore. For $\beta$ a pure complex number
the operator $h^{(PT)}$ is invariant under the operation of complex conjugation followed by the transposition $w_1\leftrightarrow w_2$.
Thus the system is PT-symmetric and the energy eigenvalues are real. For comparison, let us work out the explicit form of such a PT-symmetric operator.

The unitary equivalent operator
$ H^{(PT)} \ = \ \tilde g\ h^{(PT)}\ \tilde g^{-1}$,
where \hfill\break
$\tilde g \ = \ e^{-\alpha\,{(w_1+w_2)}^2}\,\Psi_0(w_1,w_2)$,
written in Cartesian variables $(x,y)$ reads
\[ H^{(PT)} \ = \ H + \frac{\beta}{2}\bigg(  \frac{1}{x}\partial_x - \frac{1}{y}\partial_y\bigg) + V_N. \]
Here $H\ = \ \Delta_2+\frac{a_1}{x^2}+\frac{a_2}{y^2}-a_3(x^2+y^2) +a_4$
is the original operator we started with (systems $[211]$) and  {\small
\[V_N \ = \ -\frac{\alpha\,(x^2+y^2)}{16}\bigg[ \  4\,(\,6-k_1-k_2) +16\,N-8 k_3(x^2+y^2) +\alpha{(x^2+y^2)}^2    \bigg]\ ,\]}
$a_1=(-k_1^2+1)/4$, $a_2=(-k_2^2+1)/4$, $a_3=k_3^2$. Unlike $H$, the operator $H^{(PT)}$ does not admit separation of variables.

\item System $[22]$:

This system is $R$-separable in 2 sets of coordinates
 \bea\label{hyperbolic}
 &(a)&\ {\rm  Polar:}\quad\quad \ x=r\cos\theta,\ y=r\sin\theta,\\
 &(b)&\ {\rm Hyperbolic:}\quad\  x=\frac{u^2+v^2+u^2v^2}{2uv},\ y= \frac{u^2+v^2-u^2v^2}{2uv}.\nonumber\eea

Strictly, this system is neither ES nor QES, though of a simple type where the polynomial solutions are explicit!

The polynomial variables are $ w_1\ =\ x^2+y^2$, $ w_2\ =\ ({x-i\,y})/({x+i\,y})$.
With the gauge factor
$ \Psi_0\ =\ e^{(k_1\,w_1+k_2\,w_2)}\,(w_1 w_2)^{-k_3} $,
$\Psi(x,y)=\Psi_0\ P(w_1,w_2)$, and $ a_1\,=\,4\,k_2(2n_2-2k_3+1)$, $a_2\,=\,4\,k_2^2,\ a_3=-4k_1(2n_1-2k_3+1)$, $ a_4\,=\,4\,k_1^2$,
we have {\small
\[h^{(ES)}\,P\equiv \left[4w_1^2\partial_{w_1}^2-4w_2^2\partial_{w_2}^2+4w_1(2k_1w_1-2k_3+1)
\partial_{w_1}\right.\]
\[ \left.-4w_2(2k_2w_2-2k_3+1)\partial_{w_2}+8(-k_1n_1w_1+k_2n_2w_2)\right]P=0.\]}
Here, $n_1,\,n_2$ are nonnegative integers. The operator $h^{(ES)}$ maps polynomials of order $n_1$ in $w_1$ and $n_2$ in $w_2$ without increasing the order.
A  set of eigenfunctions of  $L_1$ is
\[ f_n(w_1,w_2)=(w_1w_2)^n\times\]\[{}_1F_{1}\left(\ba{c} n-n_1\\1+2(n-k_3)\ea ;-2k_1w_1\right){}_1F_{1}\left(\ba{c} n-n_2\\1+2(n-k_3)\ea ;-2k_2w_2\right),\]
where $0\le n\le n_1,\ n_2$  and $n$ is an integer.

The basis of conformal symmetries is given by $\{L_1,\,L_2,\,h^{(ES)}\}$ {\small
\bea L_1&=&-4w_2^2\partial_{w_2}^2-4w_2(2k_2w_2-2k_3+1)\partial_{w_2}-4\left(-k_2n_2w_2\right.\nonumber\\
&+&\left.(k_2-k_3)^2+2k_2n_2+k_2\right),\nonumber\\
L_2&=&\frac{4w_1}{w_2}\partial_{w_1}^2+8\partial_{w_1w_2}+\frac{4w_2}{w_1}\partial_{w_2}^2+\frac{8}{w_2}(k_1w_1+k_2w_2-2k_3)\partial_{w_1}\nonumber\\
&+&\frac{8}{w_1}(k_1w_1+k_2w_2-2k_3)\partial_{w_2}-\frac{8k_3}{w_1w_2}(2k_1w_1+2k_2w_2-2k_3-1).\nonumber\eea}
Here, $L_1$ maps polynomials to polynomials, subject to the constraints listed above, but $L_2$ does not.
This implies that polynomial solutions can be realized in polar coordinates (products of associated Laguerre polynomials),
but not in hyperbolic coordinates (products of double confluent Heun functions).

For $\tau=\exp(-2\,i\,\theta)$,
 $P(r,\tau)=S(r)T(\tau)$ the separation equations in polar coordinates are
 \[r^2\,S''(r)+r(4k_1r^2+2k_3-4m-1)\,S'(r)+8\,k_1\,m\,r^2\,S(r)\ =\ \Lambda\, S(r),\]
 \[ 4\,\tau^2\,T''(\tau)+4\tau(2k_2\tau+k_3-2m)\,T'(\tau)-8\,k_2\,m\,\tau\, T(\tau)\ =\ \Lambda \,T(\tau)\ , \]
 where $\Lambda
 $ is the separation parameter. These equations  have $m+1$ eigenvalues $\Lambda_j$ on the spaces of polynomials of maximum order
 $m$ in $r^2$ and $\tau$,
 respectively.

\item System $[31]$:

The system is exactly-solvable and $R$-separable in 2 sets of coordinates
\bea\label{parabolic} &(a)&\ {\rm Cartesian:}\ \ \
x,\ y,\\
 &(b)&\ {\rm  Parabolic:}\quad \ x= \xi^2-\eta^2,\ y=2\,\xi\,\eta\ .\nonumber\eea

The polynomial variables are
$w_1=x,\ w_2=y^2$.
The gauge factor is given by
\[ \Psi_0\ =\ e^{(k_3(w_1^2+\frac{w_2}{2})+k_2\,w_1)}\, w_2^{k_4},\quad \Psi(x,y)\,=\,\Psi_0\, P(w_1,w_2)\ .\]
With $ a_2\,=\,4\,k_2\,k_3$, $  a_3\,=\,-k_3^2$, $a_4\,=\,-2\,k_4\,(2\,k_4-1)$ ,
we arrive at  {\small
\bea   h^{(ES)}\,P\equiv  [ \partial_{w_1}^2+4\,w_2\,\partial_{w_2}^2+
(2\,k_2+4\,k_3\,w_1)\partial_{w_1}+(8\,k_4+2+4\,k_3\,w_2)\partial_{w_2} \nonumber  \eea
\bea\label{P31} +k_2^2+4\,k_3\,k_4+3\,k_3+a_1\,]\,P\ =\ 0  \  . \eea }
In variables $w_1$ and $w_2$ the operator $h^{(ES)}$ maps $2$-variate polynomials into polynomials without increasing the order and is formally self-adjoint with respect to the inner product
 \[ \langle P_1 , P_2 \rangle \equiv \int \int \, dw_1\,dw_2\, P_1(w_1,w_2)\,\overline{P_2}(w_1,w_2)\,{|\Psi_0|}^2\,w_2^{-\frac{1}{2}} \ .  \]
Boundaries of the configuration space (domain), $- \infty < w_1 < \infty\ ,\,0 \leq w_2$ are determined by the zeros and singularities of $\Psi_0$. Square-integrability demands $k_4\geq \frac{1}{4}$,\ $k_3< 0$.
In terms of $gl_3$ generators ${\cal J}$'s (\ref{generators}) the equation (\ref{P31}) reads
 \[  h^{(ES)}\,P=\left[\   {\cal J}^-_{1}{\cal J}^-_{1}   +   2\,k_2\,{\cal J}^-_{1} +4\,k_3\,{\cal J}^0_{11}  + 4\,{\cal J}^0_{22}{\cal J}^-_{2} \right. \]
\[\left.  + (8\,k_4+2)\,{\cal J}^-_{2}  +  4\,k_3\,{\cal J}^0_{22}  + a_1 +k_2^2+3\,k_3+4\,k_3\,k_4  \    \right]P=0 \  .\]
If $N$ is a non-negative integer with the constraint
\[ a_1+ k_3\,(\,4\,N+4\,k_4+3)+k_2^2\ =\ 0 \ , \]
then the operator $h^{(ES)}$ admits a basis of $2$-variate polynomial solutions of highest total order $N$ in $w_1,w_2$.

The basis of symmetries is given by the operators $\{L_1,\,L_2,\,h^{(ES)}\}$
\bea L_1&=&\partial_{w_1}^2+ (4\,k_3\,w_1+2\,k_2)\partial_{w_1}+(k_2^2+2\,k_3),\nonumber\\
L_2&=&(-k_3\,w_2-2\,k_4-\frac12)\partial_{w_1}+(-2\,k_2\,w_2+8\,k_4\,w_1+2\,w_1)\partial_{w_2}\nonumber
\\
&-&2\,w_2\,\partial_{w_1w_2}+\,w_1\,w_2\partial_{w_2}^2-\frac12 k_2(4\,k_4+1).\nonumber\eea
These operators map polynomials to polynomials without increasing  degree.

The separation equations in Cartesian  coordinates are exactly solvable (products of Hermite and associated Laguerre polynomials);
for parabolic coordinates they are QES (products of biconfluent Heun polynomials). Indeed, setting
 $P(u,v)=U(u)V(v)$ where $u=\xi^2,v=\eta^2$, and  with the constraint mentioned above we find the separation equation
 \[2\,u\,U''(u)+\left(2k_3u^2+2k_2u+4k_4+1)\right)\,U'(u)-2\,k_3\,N\,u\,U(u)\,=\, \Lambda \, U(u)\  ,\]
 with an exactly similar equation for $V(v)$,
 where $\Lambda
 $ is the separation constant. These equations have $N+1$ eigenvalues $\Lambda_j$ on the space of polynomials of maximum
 order $N$ in $\xi^2$ and $\eta^2$,
 respectively.

\item System $[4]$:

Similar to $[22]$, this system is neither ES nor QES. It is $R$-separable only in the family of semi-hyperbolic coordinates, a representative of which is
\be \label{semihyp} x=-(w-u)^2 + i\,(w + u),\quad y =-i\,(w-u)^2 + (w + u)  \ .\ee
The polynomial variables are
\[ w_1 \,=\, \frac{1}{4}(y-i\,x-i\sqrt{2(x-i\,y)}\,) \ ,\ w_2 \,=\, \frac{1}{4}(y-i\,x+i\sqrt{2(x-i\,y)}\,)  \ .  \]
Setting $a_4\,=\,-k_4^2$ and with gauge factor  as {\small
\[ \Psi_0(w_1,w_2) =\]
\[\exp \left[\frac{1}{6k_4^3}\left\{ -32k_4^4(w_1^3+w_2^3)+12ik_4^2a_3(w_1^2+w_2^2)-3
(k_4^2a_2+a_3^2)(w_1+w_2)\right\}\right],\]}
we obtain the equation
\[h^{(ES)}\,P\equiv \left[ -\left(\partial_{w_1}^2+\frac{(-32\,k_4^4\,w_1^2+8\,i\,k_4^2\,a_3\,w_1-k_4^2\,a_2-a_3^2)}{k_4^3}\partial_{w_1}\right)\right.\]
\[+\left(\partial_{w_2}^2+\frac{(-32\,k_4^4\,w_2^2+8\,i\,k_4^2\,a_3\,w_2-k_4^2\,a_2-a_3^2)}{k_4^3}\partial_{w_2}\right)\]
\[\left.-\frac{4i}{k_4^4}(w_1-w_2)(8\,i\,k_4^5-2\,k_4^4\,a_1-k_4^2\,a_2\,a_3-a_3^2)\right]\,P \ = \ 0   \  .\]
For a non-negative integer $N=0,1,2,...$, the operator $h^{(ES)}$ maps polynomials of order $\le N$ in both $w_1$ and $w_2$ when
\be \label{constraint5} a_1+4\,i\,k_4\,N+\frac{a_2\,a_3}{2\,k_4^2}+\frac{a_3^2}{2\,k_4^4}-4\,i\,k_4\ =\ 0  \ . \ee

Note that
$h^{(ES)}\,=Y_1-Y_2$ where
\[ Y_s=-\left(\partial_{w_s}^2+\frac{(-32k_4^4w_s^2+8ik_4^2a_3w_s-k_4^2a_2-a_3^2)}{k_4^3}\partial_{w_1}\right)+32\,k_4\,N\,w_s,\]
for $s=1,2$.
Now let $f$ be a polynomial eigenfunction of $Y_s$ of order $N$ (which must exist since the space of polynomials of order $\le N$ in $w_s$ is
invariant under $Y_s$):
$Y_s\,f(w_s)\ =\ \lambda\, f(w_s)$, $s=1,2$,
and set $P=f(w_1)f(w_2)$. Then
\[ h^{(ES)}\,P=(Y_1-Y_2)f(w_1)f_(w_2)=(\lambda -\lambda)f(w_1)f(w_2)=0,\]
There are $N+1$  eigenfunctions, (triconfluent Heun polynomials).

The basis of conformal symmetries consist of the operators $\{ L_1,\,L_2,\,h^{(ES)}\}$,
\bea L_1&=&-\frac1{8(w_1-w_2)^3k_4^5}\left(16w_1^3k_4^6-16k_4^6w_1^2w_2-16k_4^6w_1w_2^2\right.\nonumber\\
&+&\left.16w_2^3k_4^6(4ia_3k_4^4w_1^2+8ia_3k_4^4w_1w_2-4ia_3k_4^4w_2^2+k_4^5\right)\partial_{w_1}\nonumber\\
&+&\frac{1}{8(w_1-w_2)^3k_4^5}\left(16w_1^3k_4^6-16k_4^6w_1^2w_2-16k_4^6w_1w_2^2+16w_2^3k_4^6\right.\nonumber\\
&-&\left. 4ia_3k_4^4w_1^2+8ia_3k_4^4w_1w_2-4ia_3k_4^4w_2^2+k_4^5\right)\partial_{w_2}\nonumber\\
&+&\frac{1}{16(w_1-w_2)^2}\partial_{w_1}^2-\frac{1}{8(w_1-w_2)^2}\partial_{w_1w_2}+\frac{1}{16(w_1-w_2)^2}\partial_{w_2}^2
-\frac{a_3^2}{k_4^2},\nonumber\\
 L_2&=&\frac{w_2(-32w_1^2k_4^4+8ik_4^2a_3w_1-k_4^2a_2-a_3^2)}{(w_1-w_2)k_4^3}\partial_{w_1}-\nonumber\\
 && \frac{w_1(-32w_2^2k_4^4+8ik_4^2a_3w_2-k_4^2a_2-a_3^2)}{(w_1-w_2)k_4^3}\partial_{w_2}+\frac{w_2}{(w_1-w_2)}\partial_{w_1}^2\nonumber\\
 &-&\frac{w_1}{(w_1-w_2)}\partial_{w_2}^2-\frac{1}{4k_4^{6}}(16ik_4^{5}a_3+(a_3^4+k_4^4a_2^2+2k_4^2a_2a_3^2).\nonumber\eea
Neither $L_1$ nor $L_2$ maps polynomials to polynomials.
However, assuming constraint (\ref{constraint5}), $L_2$ maps polynomial solutions of $h^{(ES)}\,P=0$ with order $\le N$ to solutions.
Indeed
\[ L_2=J_1+\frac{2k_4^2a_2a_3^2+k_4^4a_2^2+16ik_4^5a_3+a_3^4}{4k_4^6}+\frac{w_1}{w_2-w_1}\,h^{(ES)}\   .\]

\item System $[0]$:

This system separates in a family of Cartesian coordinates. The polynomial variables are
$w_1=x-\frac{a_2}{2a_4}$, $w_2=y-\frac{a_3}{2a_4}$.
Note that in this case the polynomial coordinates depend on the parameters in the potential.
With
\[ a_2=4\,k_4\,k_2,\  a_3=4\,k_4\,k_3,\ a_4=-4\,k_4^2,\quad
 \Psi_0=\exp\left[k_4(w_1^2+w_2^2)\right]\]
 and $\Psi(x,y)=\Psi_0(w_1,w_2)\,P(w_1,w_2)$, we have {\small
 \[h^{(ES)}\,P\equiv \left(  \partial_{w_1}^2+
\partial_{w_2}^2  + 4k_4w_1\partial_{w_1}+4k_4w_2\partial_{w_2}+k_2^2+k_3^2+a_1+4k_4\right)P=0.\]}
The operator $h^{(ES)}$ is exactly-solvable, it maps polynomials into polynomials without increasing the order and is formally self-adjoint with respect to the inner product
 \[ \langle P_1 , P_2 \rangle \equiv \int \int \, dw_1\,dw_2\, P_1(w_1,w_2)\,\overline{P_2}(w_1,w_2)\,{|\Psi_0|}^2 \ .  \]
The boundaries of the configuration space (domain) are $- \infty < w_{1,2} < \infty$. Square-integrability demands $k_4< 0$.
In terms of the $gl_3$ generators ${\cal J}$`s, (\ref{generators}): {\small
 \[  h^{(ES)}\,P=\left[\ {\cal J}^-_{1}{\cal J}^-_{1} + {\cal J}^-_{2}{\cal J}^-_{2} + 4\,k_4\,({\cal J}^0_{1}+{\cal J}^0_{2} )
 + k_2^2+k_3^2+a_1+4k_4  \    \right]P=0 \  .\]}
If $4\,k_4\,N + k_2^2+k_3^2+a_1+4k_4\ =\ 0$,
then $h^{(ES)}$ admits a basis of polynomial solutions of highest total order $N$ in $w_1,w_2$.
The basis of symmetries is given by $\{L_1,\,L_2,\,h^{(ES)}\}$:
\bea L_1& =& 4k_4w_1\partial_{w_1}+\partial_{w_1}^2+k_2^2+2k_4,\nonumber\\
 L_2&=&2k_4w_2\partial_{w_1}+2k_4w_1\partial_{w_2}+\partial_{w_1w_2}+k_2k_3.\nonumber\eea

\item System $(1)$:

This exceptional conformally superintegrable system is not exactly or quasi-exactly solvable. However, if we assume factorizable solutions
of the form $\Psi(x,y)=\exp(\lambda {\bar z})\psi(z)$ for $z=x+iy,{\bar z}=x-iy$, then for $\Psi=\exp(\lambda {\bar z})\Theta(z)P(z)$
with
$ \Theta\,=\,\exp\left[\frac{1}{4\lambda}\bigg(\frac{a_1}{z}-\frac{a_3}{2z^2}+\frac{a_4}{3z^3}\bigg)\right]$,
we have $(4\lambda \frac{d}{dz}+a_2)P(z)=0$. The operator  $(4\lambda \frac{d}{dz}+a_2)$ takes
polynomials in $z$ to polynomials, without increasing the degree. For $4\,\lambda\,N+a_2=0$ the operator $(4\lambda \frac{d}{dz}+a_2)$ admits a basis of polynomial solutions of highest total order $N$ in $z$.

The basis of symmetries is $\{L_1,L_2,H\}$:
\bea L_1 &=& \partial_x+i\partial_y  \ ,\nonumber\\
 L_2&=&\{x\partial_y-y\partial_x,\partial_x+i\partial_y\} +i\left(\frac{2a_1}{(x+iy)}-\frac{3a_3}{2(x+iy)^2}+\frac{4a_4}{3(x+iy)^3}\right).\nonumber\eea

\item System $(2)$:
As with the preceding case, this exceptional conformally superintegrable system is neither exactly nor quasi-exactly solvable.
However, if we assume solutions
of the form $\Psi(x,y)=\exp(\lambda {\bar z})\psi(z)$ for $z=x+iy,{\bar z}=x-iy$, then for $\Psi=\exp(\lambda {\bar z})\Theta(z)P(z)$
with
$\Theta\ =\ \exp\left[-\frac{1}{4\lambda}(\frac{a_2\,z^2}{2}+\frac{a_3\,z^3}{3}+\frac{a_4\,z^4}{4})\right]$
we have $(4\lambda \frac{d}{dz}+a_1)P(z)=0$. The operator  $(4\lambda \frac{d}{dz}+a_1)$ takes
polynomials in $z$ to polynomials, without increasing the degree.

In this case the basis of symmetries is $\{L_1,\,L_2,\,H\}$
{\small
\bea L_1 & = &\partial_x+i\partial_y\ ,\nonumber\\ L_2&=&\{x\partial_y-y\partial_x,\partial_x+i\partial_y\}
  +i\left(\frac{a_2(x+iy)^2}{2}+\frac{2\,a_3(x+iy)^3}{3}
 +\frac{3\,a_4(x+iy)^4}{4}\right).\nonumber\eea }

\end{enumerate}

\section{Degenerate potentials}

In this section we consider Laplace superintegrable systems with two-parameter potentials. In total there are six degenerate potentials, listed in Table \ref{Tab2}. To distinguish them from the non-degenerate ones hereafter the polynomial variables $(w_1,\,w_2)$ will be denoted as $(u_1,\,u_2)$. The two-parameter potentials can all be obtained as parameter restrictions of the previous four-parameter potentials, though they have additional symmetry.

\begin{enumerate}
\item{\bf System A }

This system is ES,  and  $R$-separable in two coordinate systems:
\bea\label{spherical2} &(a)&\ {\rm Spherical:}\
u_1 = \sin \theta \cos \phi,\ u_2 = \sin \theta \sin \phi,\\ \label{elliptic2}
 &(b)&\ {\rm  Elliptic:}\quad\  u_1^ 2 = \frac{(c\,u - 1)(c\,v - 1)}{(1 - c)},\
 u_2^2=\frac{c\,(u-1)(v-1)}{(c-1)},\nonumber \\
& &\qquad c \ {\rm is\ a\ parameter}\ \ne 0,1,\eea

  The polynomial variables are $u_1=\frac{2x}{x^2+y^2+1}, u_2=\frac{2y}{x^2+y^2+1}$.
 With
 $  a_3=-2\,k_3\,(2\,k_3-1)$,  $ \Psi_0=(1-u_1^2-u_2^2)^{k_3}$,  and $\Psi(u_1,\,u_2)=\Psi_0\ P(u_1,u_2)$, we have
 \[h^{(ES)}\,P\equiv  \left[ (1-u_1^2)\partial_{u_1}^2+(1-u_2^2)\partial_{u_2}^2-2\,u_1\,u_2\partial_{u_1u_2}^2\right.\]
\[\left. -u_1\left(2+4\,k_3\right)\partial_{u_1}-u_2\left(2+4\,k_3\right)\partial_{u_2}
 -(2k_3(2k_3+1)+a_4)\right]P=0.\]

The operator $h^{(ES)}$ acting on $P(u_1,\,u_2)$ maps polynomials into polynomials without increasing the order and is formally self-adjoint with respect to the inner product
 \[ \langle P_1 , P_2 \rangle \equiv \int \int \, du_1\,du_2\, P_1(u_1,u_2)\,\overline{P_2}(u_1,u_2)\,{|\Psi_0|}^2\,{(1-u_1^2-u_2^2)}^{-\frac{1}{2}} \ .  \]
The boundaries of the configuration space (domain) $u_1\geq0,\,u_2\geq0$ and $u_1^2+u_2^2\leq1$ corresponds to the zeros of $\Psi_0$. Square-integrability demands $k_3\geq \frac{1}{4}$.

In terms of generators ${\cal J}$`s, (\ref{generators}),
\[h^{(ES)}\,P \ =\  \left[  {\cal J}^-_{1}{\cal J}^-_{1}+{\cal J}^-_{2}{\cal J}^-_{2} - ({\cal J}^0_{1}{\cal J}^0_{1}+{\cal J}^0_{2}{\cal J}^0_{2}) -(1+4\,k_3)\,({\cal J}^0_{1}+{\cal J}^0_{2})\right.\]
\[ \left. -2{\cal J}^0_{1}{\cal J}^0_{2} -(2k_3(2k_3+1)+a_4)
\right]P=0.\]
There are infinitely many finite-dimensional invariant subspaces
${\cal P}_{M}^{(2)} \ =\ \langle  u_1^{p_1}\,u_2^{p_2}\mid 0\leq p_1+p_2\leq  M \rangle$
where $M=0,1,2...$,
%which form flag:
%\[  {\cal P}_0^{(2)}\subset {\cal P}_1^{(2)} \subset {\cal P}_2^{(2)}...\subset {\cal P}_M^{(2)} \subset ...\,{\cal P} \ . \]
provided
\begin{equation}
(M+2\,k_3)(M+2\,k_3+1)+a_4  \ = \ 0   \ .
\label{polyconstraintA}
\end{equation}

Again, by adding the raising operator
\[ {\cal J}^+(M) = (u_1+u_2)(u_1\partial_{u_1}+u_2\partial_{u_2}-M)   \ ,  \]
the exactly-solvable operator $h^{(ES)}$ that annihilates $P(u_1,\,u_2)$ becomes quasi-exactly-
solvable $ h^{(QES)}=h^{(ES)} + \alpha\,{\cal J}^+(M)$,
where $\alpha$ is a real parameter. The operator $h^{(QES)}$ has a single invariant subspace in $2$-variate polynomials. By adding the term
\[ I = {\cal J}^-_{1}-{\cal J}^-_{2},   \quad
 h^{(PT)}=h^{(ES)} + \alpha\,{\cal J}^+(M) + \beta\,I \ , \]
with $\beta$ a parameter, the operator $h^{(PT)}$ is quasi-exactly-solvable. However, it is not formally self-adjoint due to the boundary terms. For $\beta$ a pure complex number, $h^{(PT)}$ is invariant under the operation of complex conjugation followed by the transposition $u_1\leftrightarrow u_2$. Thus the system is PT-symmetric and the energy eigenvalues are real.

For conformal symmetries we can take the basis $\{L_1=J^2, L_2,L_3,H\}$,
\[ J=u_1\partial_{u_2}-u_2\partial_{u_1},\
 L_2=-\frac14(u_1^2+u_2^2-1)\partial_{u_1}^2-(\frac14+k_3)u_1\partial_{u_1},\]
 \[L_3=-2(u_1^2+u_2^2-1)\partial_{u_1u_2}-(1+4k_3)(u_2\partial_{u_1}+u_1\partial_{u_2}).\]
 Solutions $P(u_1,u_2)$ are separable in spherical coordinates (\ref{spherical2}). They are products of Jacobi polynomials corresponding to eigenfunctions of $L_1$; those separable in
 elliptic coordinates (\ref{elliptic2}) are eigenfunctions of $-\frac14 L_1+(r-1)L_2$ and take the form of products of Heun polynomials. Indeed, in elliptic coordinates the conformal symmetry operator
 \[ K \ \equiv \ -\frac14 \, L_1+(r-1)\,L_2 -\frac14 (r\,v-1)\,H\]
 \[=v(v-1)(rv-1)\partial_{vv}
 +\left(\frac12-(1+r)(2\,k_3+1)v+r(2\,k_3+\frac32) v^2\right)\partial_v\]
\[ + \frac14 r  (4\,k_3^2+2\,k_3+a_4)v-\frac14 (4\,k_3^2+2\,k_3+a_4)\ ,  \]
is QES. For
$a_4+2\,(N+k_3)(2\,N+2\,k_3+1)\ = \ 0$,
(i.e., $M=2N$) maps the space of polynomials in $v$ with maximum order $N$ into itself.

 \item {\bf System B}

This system is $R$-separable in four coordinate systems
\bea\label{Cart1} &(a)&\ {\rm Cartesian:}\
x,\ y,\\
&(b)&\ {\rm Polar}:\
x=r\cos\theta,\ y=r\sin\theta.\nonumber\\
&(c)&\ {\rm Parabolic}: x=\xi\,\nu,\ y=\frac12(\xi^2-\nu^2),\nonumber\\
 &(d)&\ {\rm  Elliptic:}\quad\  x^ 2 = c^2( u - 1)( v - 1),\
 y^2=-c^2\,u\,v,\nonumber \\
& &\qquad c  \ {\rm is\ a\ parameter}\ \ne 0\ .\nonumber \eea

The polynomial variables are $u_1,\,u_2$ where
 $x^2\, =\, u_1\,u_2$ , $ y\, =\, \frac12(u_1-u_2)$.
Setting
\[a_4=A^2,\ a_1=-(N+2)(N+\frac{1}{2}),\ \Psi_0=\exp[\frac{i\,A}{2}(u_1+u_2)](u_1\,u_2)^{-N/2-1/4},\]
and $\Psi=\Psi_0(u_1,u_2)\,P(u_1,u_2)$, we have
 \[\frac14(u_1+u_2)\ h^{(ES)}\,P\equiv  \]
\[ \left[ (u_1\partial_{u_1}^2+u_2\partial_{u_2}^2+i\,(A\,u_1+i\,N)\partial_{u_1}+i\,(A\,u_2+i\,N)
\partial_{u_2}-i\,A\,N   \right]\ P =0 \ .\]
In terms of generators ${\cal J}$`s (\ref{generators})
\[ \left[  {\cal J}^0_{1}{\cal J}^-_{1}+{\cal J}^0_{2}{\cal J}^-_{2} +i\,A\,({\cal J}^0_{1}+{\cal J}^0_{2}) -N \,({\cal J}^-_{1}+{\cal J}^-_{2})-   i\,A\,N
\right]\ P=0.\]
This equation admits a basis of polynomial solutions of highest total order $N$ in $u_1,u_2$, provided $N$ is a non-negative integer. In this case the functions $\{ \Psi \}$ are not square integrable in the domain $u_1,\,u_2\in [0,\,\infty).$

The generating symmetries are $\{P_y, L_1,L_2,H\}$ where
\[ P_y=\partial_y,  \quad
 L_1 = x^2\partial_{yy}-2xy\partial_{xy}+y^2\partial_{xx}-x\partial_x-y\partial_y+\frac{a_1y^2}{x^2}\ ,\]
 \[ L_2= x\partial_{xy}-y\partial_{xx}+\frac12\partial_y-\frac{a_1y}{x^2}\ .\]
Solutions separable in Cartesian coordinates are eigenfunctions of $P_y^2$, and  $P$ is a product of a Bessel function (not a polynomial) and a monomial.
Those separable in polar coordinates are eigenfunctions of $L_1$, and $P$ is a product of a Bessel function (not a polynomial) and an associated Laguerre function.
Those separable in parabolic coordinates are eigenfunctions of $L_2$; $P$ is a product of two confluent hypergeometric functions.
Those separable in elliptic coordinates are eigenfunctions of $L_1-c^2P_y^2$; $P$ is a product of two spheroidal wave functions.

 \item{\bf System C}

This system, the harmonic oscillator, is $R$-separable in coordinate systems
\bea\label{Cart2} &(a)&\ {\rm Cartesian:}\
x,\ y,\\
&(b)&\ {\rm Polar}:\
x=r\cos\theta,\ y=r\sin\theta.\nonumber\\
&(c)&\ {\rm Hyperbolic}: x=\frac{U^2+V^2+U^2V^2}{2UV},\ y=i\,\frac{U^2+V^2-U^2V^2}{2UV},\nonumber\\
 &(d)&\ {\rm  Elliptic:}\quad\  x^ 2 = c^2\,( u - 1)( v - 1)\ ,\
 y^2=-c^2\,u\,v  \ ,\nonumber \\
& &\qquad c \ {\rm a\ parameter}\ \ne 0  \ .  \nonumber \eea

The polynomial variables are
$u_1 = x^2$,  $u_2=y^2$ .
Because of the $Z_2$ symmetry, the null space of $H$ splits into even and odd functions. Correspondingly, there are two possible gauge transformations. Putting $a_4=-k_4^2$, the first gauge transformations is
$ \Psi_{0,1}=\exp[k_4(x^2+y^2)]$ .
With
$\Psi(x,y)\ =\ \Psi_{0,1}\,P(u_1,u_2)$
we have

\[h^{(ES)}\,P\equiv\] \[\left[(2+8k_4u_1)\partial_{u_1}+(2+8k_4u_2)\partial_{u_2}+4u_2\partial_{u_2}^2+4u_1\partial_{u_1}^2+(4k_4+a_1)\right]\,P=0\ .\]
If
$a_1+4\,k_4\,(2\,N+1)\ =\ 0$
and $N$ is a nonnegative integer then $h^{(ES)}$ leaves invariant the space of polynomials in $u_1,u_2$ of maximum order $N$.
The second transformation is
\[  \ \Psi_{0,2}=\exp[k_4(x^2+y^2)]\,x\,y\ , \qquad  \Psi(x,y)=\Psi_{0,2}\,P(u_1,u_2)\ ,  \]
and the corresponding equation takes the form
\[h^{(ES)}\,P   \equiv\] {\small
\[ \left[(6+8k_4u_1)\partial_{u_1}+(6+8k_4u_2)\partial_{u_2}+4u_2\partial_{u_2}^2+4u_1\partial_{u_1}^2+(12k_4+a_1)\right]\,P=0\ .\]}
If
$a_1+4\,k_4\,(2\,M+3)\ =\ 0$
and $M$ is a nonnegative integer then $h^{(ES)}$ also leaves invariant the space of polynomials in $u_1,u_2$ of maximum order $M$.
For both $\Psi_{0,1}$ and $\Psi_{0,2}$, the corresponding operator $h^{(ES)}$ is exactly-solvable and is formally self-adjoint with respect to the inner product
 \[ \langle P_1 , P_2 \rangle \equiv \int \int \, dw_1\,dw_2\, P_1(w_1,w_2)\,\overline{P_2}(w_1,w_2)\,|\Psi_{0,1(2)}|^2 \ .  \]
The configuration space is $w_1\geq0,\,w_2\geq0$\,.

The operators $\{L,L_1,L_2,H\}$
\[ L=x\partial_y-y\partial_x\,,\  L_1=\partial_{x}^2-4k_4^2x^2\,,\  L_2= \partial_{xy}-4\,k_4^2\,x\,y\ .\]
form a basis of symmetries. Solutions separable in Cartesian coordinates are eigenfunctions of $L_1$, $P$ is a product of Hermite functions.
Those separable in
polar coordinates are eigenfunctions of $L^2$, and $P$ is an associated Laguerre function times a monomial. Solutions separable in
hyperbolic coordinates are eigenfunctions of $L_1+iL_2+\frac12 L^2$,
and $P$ is a product of Whittaker functions. Those solutions separable in elliptic coordinates are eigenfunctions of $L_1+c^{-2}L^2$ and $P$ is a product of confluent Heun functions.
The polynomial spaces do not admit separated solutions  in hyperbolic coordinates.

\item{\bf System D}

This system is $R$-separable in two coordinate systems.
\bea\label{Cart3} &(a)&\ {\rm Cartesian:}\
x,\ y,\\
&(b)&\ {\rm Parabolic}: y=\xi\nu,\ x=\frac12(\xi^2-\nu^2).\nonumber\eea
In this case there are no polynomial invariant subspaces. The generating symmetries of $H$ are $\{P_y, L_1,L_2,H\}$ where
 \[ P_y=\partial_y\ ,\ L_1=\frac12 \{x\partial_y-y\partial_x,\partial_y\}+\frac{a_2}{4}y^2\ ,\ L_2=\partial_{xy}-\frac{a_2}{2}y\ .\]
Solutions separable in Cartesian coordinates are eigenfunctions of $P_y^2$, they are given by the product of an Airy function times an exponential. Those separable in parabolic coordinates are eigenfunctions of $L_1$, they factorize as the product of triconfluent Heun functions.

\item{\bf System E}

This  is $R$-separable in two coordinate systems.
\bea\label{polar} &(a)&\ {\rm polar:}\
x=r\cos\theta,\ y=r\sin\theta,\\
&(b)&\ {\rm hyperbolic:}\ \nonumber\\
&&x=\frac{U^2+V^2+U^2V^2}{2UV},\ y=i\frac{U^2+V^2-U^2V^2}{2UV}.\nonumber\eea
As for the system $D$, in this case there are no polynomials invariant subspaces. The generating symmetries of $H$ are $\{P_+, L_1,L_2,H\}$ where
 \[ P_+=\partial_x+i\partial_y,\ L_1=\frac12\{M,P_+\}-i\frac{a_1}{(x+iy)}\ ,\
  L_2=M^2+a_1\frac{(x-iy)}{(x+iy)}\ ,  \]
with $M=x\partial_y-y\partial_x$.
Solutions, product of Bessel functions, separable in polar coordinates are eigenfunctions of $L_2$.
Those separable in hyperbolic coordinates are products of doubleconfluent Heun functions corresponding to the eigenfunctions of $L_2+P_+^2$ \ .

\item{\bf System F}
 This system is $R$-separable in two coordinate systems.
\bea\label{Cart4} &(a)&\ {\rm Cartesian:}\
x,\ y,\\
&(b)&\ {\rm semi-hyperbolic:}\ \nonumber\\
&&x=i(w-u)^2+2i(w+u),\ y=-(w-u)^2+2(w+u).\nonumber\eea
The polynomial variables are $u_1=u,\,u_2=w$, the semi-hyperbolic coordinates.
Setting
$a_2=i\frac{k_1^2}{16}$ , $a_1=\frac{k_1\,k_4}{4}$,
together with the gauge factor
\[\Psi_0=\exp[k_1(u^2+w^2)+k_4(u+w)] =\exp[\frac{k_1}{32}(y-ix)^2-\frac{k_1}{4}(y+ix)+\frac{k_4}{2}(y-ix)],\]
$\Psi = \Psi_0\,P(u,w)$, we obtain
 \[8(u-w)h^{(ES)}\,P \equiv \]
\[ \left[ (2k_1u+k_4)\partial_{u}-(2k_1w+k_4)\partial_{w}+\frac12\partial_{u}^2-\frac12\partial_{w}^2\right]P=0\ .\]
In terms of generators ${\cal J}$`s (\ref{generators}),
\[ \left[ \frac{1}{2}({\cal J}^-_{u}{\cal J}^-_{u}-{\cal J}^-_{w}{\cal J}^-_{w}) + 2\,k_1\,({\cal J}^0_{u}-{\cal J}^0_{w})
+ k_4\,({\cal J}^-_{u}-{\cal J}^-_{w})  \right]\,P=0\ .\]
For any non-negative integer $N$ this equation admits a basis of polynomial solutions of highest order $N$ in $u$ and highest order $N$ in $w$.
This space of polynomial variables does not admit separated solutions in Cartesian coordinates.

The generating symmetries are $\{P_+, L_1,L_2,H\}$ where
\[ P_+=\partial_x+i\partial_y\ ,\quad \ L_1=\partial_x^2-a_2 x\ ,\]
\[ L_2=(x-iy)\partial_{xy}+ix\partial_{yy}-y\partial_{xx}+\frac12\partial_y-\frac{i}{2}\partial_x-\frac{i}{4}a_2(x-iy)^2\ .\]
Solutions separable in Cartesian coordinates are eigenfunctions of $L_1$, they are given by a product of Airy functions.
Those separable in
semi-hyperbolic coordinates are eigenfunctions of $P_+^2-4L_1-L_2$, i.e., they are products of parabolic cylinder functions.

\end{enumerate}

\subsection{Polynomial solutions and conformal St\"ackel transform }

For a given Laplace system our approach allows us to study in a unified way the whole family of equivalent Helmholtz systems at once.
To illustrate how this general method works we study in some detail the Laplace system $[211]$.
In particular, we will show that constraint (\ref{polyconstraint2}) encodes the eigenvalues of all Helmholtz systems conformally St\"ackel equivalent  to $[211]$. Moreover, the polynomial solutions are the same.

We start from the equation
\be\label{H211} H_{211}\,\Psi \equiv \ \left(\Delta_2+\frac{a_1}{x^2}+\frac{a_2}{y^2}-a_3(x^2+y^2)+a_4\right)\Psi\ =0, \ee
where $-a_4$ plays the role of the spectral parameter for the associated Helmholtz system
\be\label{E1} H_{E1}\Psi\equiv \left(\Delta_2+\frac{a_1}{x^2}+\frac{a_2}{y^2}-a_3(x^2+y^2)\right)\Psi=-a_4\Psi.\ee From (\ref{polyconstraint2}) we already know that $H_{211}$ possesses polynomials solutions for
$a_4\ = \- 4\,k_3\,(2\,N+2\,k_1+2\,k_2+1) $, where
$a_1=-2k_1(2k_1-1),\ a_2=-2k_2(2k_2-1),\ a_3=4k_3^2$ . Now, let us consider the St\"ackel equivalent system $[E16]$, \cite{Kress2001}.
The corresponding Helmholtz equation is the flat space system
\be\label{HE16} H_{E16}\,\Psi \equiv \frac{1}{x^2+y^2}\bigg(\Delta_2+\frac{a_1}{x^2}+\frac{a_2}{y^2}+{a_4}\bigg)\Psi  \ = \ a_3\,\Psi \ , \ee
where the roles of $-a_4$ and $a_3$ are interchanged. In this case $a_3$ plays the role of the spectral parameter while the other parameters $a_1,a_2$ and $a_4$ remain fixed. The operator $H_{E16}$ is formally self-adjoint with respect to the inner product
\[ \langle \Psi_1 , \Psi_2 \rangle \equiv \int \int \, (x^2+y^2)\,dx\,dy\, \Psi_1\,\overline{\Psi_2} \ .  \]
The domain is $-\infty < x < \infty$, $-\infty < y < \infty $.  Unlike the $[211]$ system, now the measure in the inner product contains the extra factor $(x^2+y^2)$. From the constraint we have $ k_3=a_4/[4(2N+2k_1+2k_2+1)$ so
polynomial solutions of the St\"ackel equivalent system $[E16]$ (\ref{HE16}) occur at quantized values of the spectral parameter $a_3$ given by
\[  a_3\ =\ \frac{a_4^2}{4{(1+2(k_1+k_2+N))}^2} \ . \]
Note that the dependence of the energy eigenvalue on $N$ is nonlinear.

We introduce
parabolic coordinates $u=\frac{x^2-y^2}{2}$, $v=xy $ .
In these coordinates the polynomial variables are
\[ w_1=\sqrt{u^2+v^2}+u,\  w_2=\sqrt{u^2+v^2}-u,\]
 \[\Psi(w_1,w_2)\ =\ \Psi_0(w_1,w_2)\,P(w_1,w_2) \ , \quad
 \Psi_0 = w_1^{k_1}\,w_2^{k_2}\,e^{k_3(w_1+w_2)}.\]
The operator $h^{(ES)}$ obtained from $H_{E16}$ via the gauge transformation is formally self-adjoint with respect to the inner product
 \[ \langle P_1 , P_2 \rangle \equiv \int \int \, dw_1\,dw_2\, P_1(w_1,w_2)\,\overline{P_2}(w_1,w_2)\,|\Psi_0|^2\,(w_1+w_2)\,w_1^{-\frac{1}{2}}\,w_2^{-\frac{1}{2}} \ .  \]
Again, boundaries of the configuration space (domain), $w_1\geq0,\,w_2\geq0$, are determined by zeros and singularities of $\Psi_0$.
Polynomial solutions of highest total order $N=0,1,2,...$ in $w_1,w_2$ appear when the constraint is satisfied.

As a second example, we present the Helmholtz system $S4$ on the sphere, conformally St\"ackel equivalent to [2,1,1], for which the Laplace equation takes the form
\be\label{S41}  \bigg(\frac{x^2\,y^2}{x^2+y^2}\Delta_2 -a_3\,x^2\,y^2+a_4\frac{x^2\,y^2}{x^2+y^2}+b_1\frac{x^2-y^2}{x^2+y^2}+b_2\bigg)\Psi=0. \ee
where $b_1=(a_2-a_1)/2$, $b_2=(a_2+a_1)/2$.
It defines an eigenvalue problem with spectral parameter $-b_2$. Applying the constraint equation $a_4\ = \- 4\,k_3\,(2\,N+2\,k_1+2\,k_2+1) $ to the case where $b_1,k_3,a_4$ are fixed, we find polynomial solutions for
\[ b_2=-4\frac{b_1^2}{(6-\frac{a_4}{2k_3}+4N)^2}-\frac{1}{64k_3^2}(8Nk_3-a_4+4k_3)(8Nk_3-a_4+12k_3),\]
 $N=0,1,2...\ $.

To  verify that system (\ref{S41}) is in fact a system on the 2-sphere, real on the 2-sheet hyperboloid, we note that $x,y$ are degenerate elliptic coordinates of type 2. Indeed setting
\[ s_1+is_2=-\frac{i}{xy},\ s_1-is_2=\frac{i}{4}\frac{(x^2+y^2)}{xy},\ s_3=\frac{i}{2}\frac{(y^2-x^2)}{xy},\]
we see that $s_1^2+s_2^2+s_3^2=1$ and (\ref{S41})  becomes
\[ \bigg(\Delta_2'+4b_1\frac{s_3}{\sqrt{s_1^2+s_2^2}}-\frac{a_3}{(s_1+is_2)^2}+2i\frac{a_4}{(s_1+is_2)\sqrt{s_1^2+s_2^2}}-b_2\bigg)\Psi=0,\]
where $\Delta_2'$ is the Laplace-Beltrami operator on the complex 2-sphere.

\subsubsection*{St\"ackel equivalent classical systems}

Some discussion of the relationship between Laplace-type and Helmholtz-type  classical superintegrable systems is in order.
It is well known that trajectories of St\"ackel equivalent systems are connected by swapping the role of parameters \cite{Couplingconstant}.
We use the classical system $[211]$ to illustrate such a connection. The basic ideas apply to all superintegrable systems.
The classical system $[211]$, classical counterpart of (\ref{H211}), is described by the Hamiltonian
\be\label{H211clas} {\cal H}_{211} = p^2_x + p_y^2 +\frac{a_1}{x^2}+\frac{a_2}{y^2}-a_3(x^2+y^2) +a_4 \ . \ee
where $p_x$ and $p_y$ are the canonical momenta associated with the variables $x$ and $y$, respectively.
Consider the associated Helmholtz system ${\cal H}_{E1}=E_{E1}$ where
\[{\cal H}_{E1}=  p^2_x + p_y^2 +\frac{a_1}{x^2}+\frac{a_2}{y^2}-a_3(x^2+y^2),\]
and $E_{E1}=-a_4$. This system admits basis of 3 functionally independent 2nd order constants of the motion
$\{{\cal H}_{E1},{\cal L}_1(a_1,a_2,a_3),{\cal L}_2(a_1,a_2,a_3)\}$ where the dependence on the $a_j$ is linear. A classical trajectory can be characterized by the specific values $E_{E1},L^{(0)}_1,L^{(0)}_2$,
assumed by the constants of the motion along the trajectories. The phase space $(x,y,p_x,p_y)$ for the trajectories is
4-dimensional and the equations
\[{\cal H}_{E1}= E_{E1},\ {\cal L}_1=  L^{(0)}_1,\ {\cal L}_2=L^{(0)}_2,\]
determine 3 independent intersecting hypersurfaces in that space. The intersection is a curve on which the trajectory lies.
The intersection can be computed explicitly and takes the form
\be\label{trajectories} f(x,y,a_1,a_2,a_3,E_{E1},L^{(0)}_1,L^{(0)}_2)=0, \ee
\[p_x=g(x,y,a_1,a_2,a_3,E_{E1},L^{(0)}_1,L^{(0)}_2), \ p_y=h(x,y,a_1,a_2,a_3,E_{E1},L^{(0)}_1,L^{(0)}_2).\]

Now, the classical system $[E16]$ Hamiltonian takes the form
\be\label{HE16clas}   {\cal H}_{E16} =  \frac{1}{x^2+y^2}\bigg(p_x^2+p_y^2+\frac{a_1}{x^2}+\frac{a_2}{y^2}+{a_4}\bigg) = \ a_3\ . \ee
Under the St\"ackel transformation from $E1$ to $E16$ the constants of the motion ${\cal L}_1(a_3),{\cal L}_2(a_3)$ for $E1$ transform to
${\cal L}_1'={\cal L}_1({\cal H}_{E16})$, ${\cal L}_2'={\cal L}_2({\cal H}_{E16})$, the corresponding constants of the motion for $E16$.
Thus, for the choices $ E_{E16}=a_3$, $E_{E1}=-a_4$, ${\cal L}_1'={\cal L}_1$, ${\cal L}_2'={\cal L}_2$ on the fixed trajectory.
From this we conclude that equations (\ref{trajectories}) for the
trajectories of $E1$ are exactly the same as the equations for the trajectories of $E16$.

There are differences, of course. The coordinates $(x,y)$ have entirely different meanings for the two systems. Moreover,
changing the energy for one system corresponds to changing
a parameter in the potential function for the other. As a simple example, the 2D Kepler problem and the 2D isotropic oscillator are St\"ackel
equivalent. However, the attractive oscillator corresponds to bounded (ellipsoidal) Kepler trajectories, the repulsive oscillator to unbounded (hyperbolic) trajectories, and
the oscillator with zero force constant to parabolic Kepler trajectories.

Via a simple coordinate transformation the St\"ackel transform relates the trajectories of (\ref{HE16clas}) and (\ref{H211clas}) explicitly.
However, when looking at the dynamics there is a subtlety on the time variable. In order to explain it let us consider the
evolution of the vector position $\bf r$. For the system $E1$ we have the equation
\[  \frac{d}{dt}{\bf r} \ = \{{\cal H}_{E1},\ {\bf r}  \} =2(p_x,p_y)\ , \]
where $\{,\}$ stands for the Poisson bracket. Similarly for $E16$
\[  \frac{d}{d\tau}{\bf r} \ = \{ {\bf r},\, {\cal H}_{E16} \}=2\frac{1}{x^2+y^2}(p_x,p_y).\]
Thus,
\be \label{times} dt \ = \frac{ d\tau}{x^2(t)+y^2(t)}\,. \ee
The relation (\ref{times}) can be generalized for any pair of St\"ackel equivalent systems.

%\begin{figure}[htp]
%\begin{center}
%\includegraphics[width=2.6in,angle=0]{V211.eps}\  \includegraphics[width=2.6in,angle=0]{VE16.eps}
%\caption{Potentials $V_{211}$(left) and $V_{E16}$(right). $a_1=-4,a_2=6,a_3=-3,a_4=-8\ .$}
%\label{}
%\end{center}
%\end{figure}

%\begin{figure}[htp]
%\begin{center}
%\includegraphics[width=2.6in,angle=0]{T211.eps}\  \includegraphics[width=2.6in,angle=0]{TE16.eps}
%\caption{The trajectory for the system $[211]$(left) with energy $E_{211}=-a_4$ maps into that of $[E16]$(right) with energy $E_{E16}=a_3$. $a_1=-4,a_2=6,a_3=-3,a_4=-8\ .$}
%\label{}
%\end{center}
%\end{figure}

%\clearpage

\section{Conclusions and discussion}
Bound state eigenfunctions of time independent Schr\"odinger operators typically have the form of the ground state eigenfunction times a
polynomial. The study of these polynomial eigenspaces and their connection with special functions and orthogonal polynomials is an
important area in mathematical physics.
In this paper we have demonstrated that for the 44 2D quantum 2nd order superintegrable  systems these polynomial spaces can be classified in an
unified manner by transforming the quantum eigenvalue systems to 14 conformally superintegrable Laplace equations with potential.
Multiple 2nd order Helmholtz superintegrable systems  correspond
to a single Laplace system and determination of formal ground states and polynomial eigenspaces of the Laplace system
holds  for all Helmholtz systems
St\"ackel equivalent to it. Also we showed that the possible R-separable coordinate systems for each Laplace equation automatically separate
each associated Helmholtz system.We determined which of these 2D polynomial spaces is exactly solvable and which is quasi-exactly
solvable and we determined the possible separable
coordinate systems and associated special functions. Separation of variables in a 2D superintegrable system allowed us
to construct one-dimensional QES system.  We also showed how other 2D
QES and PT symmetric systems on constant curvature spaces can be obtained as extensions of
the solvable ones. For the associated classical mechanical 2D superintegrable systems we showed how the trajectories of
Helmholtz systems that correspond to the same
Laplace equation are related.

In other papers \cite{KMP2014,KM2014, BocherCon1,BocherCon2} we have examined the effect of contracting superintegrable systems as it applies to the symmetry algebras of the systems.
Thus we showed how the Askey scheme for orthogonal polynomials of hypergeometric type could be derived as sequences of contractions of the
symmetry  algebra of $S9$, conformally St\"ackel equivalent to $[1,1,1,1]$. Analogous procedures can be applied to the polynomial eigenspaces
of the Helmholtz systems
described here. They lead to limit relations for special functions of hypergeometric type, but also for nonhypergeometric Heun-type polynomial functions.

These ideas clearly apply to 3D 2nd order superintegrable systems on conformally flat spaces where the details are much more complicated. The systems
with nondegenerate potentials have been classified,\cite{CK2014}, but a classification of systems with degenerate 1 and 3-parameter potentials has
never been undertaken. We will address these issues in other papers.

\section{Acknowledgment}
This work was partially supported by a grant from the Simons Foundation (\# 208754 to Willard Miller, Jr) and by CONACYT grant (\# 250881 to M.A. Escobar ).


\begin{thebibliography}{99}
%-----------------------------------------------------------------------




\bibitem{BocherCon1} E. G. Kalnins, W. Miller Jr. and Eyal Subag, Laplace equations,
conformal superintegrability and B\^{o}cher contractions,
{\it Acta Polytechnica},  (to appear ), (2016). arXiv:1510.09067 [math-ph]
\bibitem{BocherCon2} E. G. Kalnins, W. Miller Jr. and Eyal Subag, B\^ocher contractions of conformally superintegrable Laplace equations,   (submitted)  	
arXiv:1512.09315 [math-ph], (2016);
B\^ocher contractions of conformally superintegrable Laplace equations: Detailed computations,
 	arXiv:1601.02876 [math-ph], (2016).


\bibitem{Bocher} B\^ocher, M., Ueber die  Reihenentwickelungender Potentialtheorie, B. G. Teubner, Leipzig (1894).

\bibitem{KKMP2011}  E.  G. Kalnins, J. M. Kress, W. Miller, Jr.  and S. Post,
Laplace-type  equations as conformal superintegrable systems,
{\it  Adv. Appl.  Math.} (2011).

\bibitem{KKM20042} Kalnins E.G., Kress J.M, and  Miller W.Jr.,
Second  order superintegrable systems in conformally flat spaces.  II: The classical 2D St\"ackel transform,
{\it J. Math. Phys.}, {\bf V.46}, 053510, (2005).

\bibitem{KKM20061} E. G. Kalnins, J. M. Kress and  W. Miller Jr.
Second order superintegrable systems in conformally flat spaces V: 2D and 3D quantum systems,
{\it J. Math. Phys.}, {\bf 47}, 093501, (2006)

\bibitem{MPW2013} W. Miller, Jr., S. Post and P. Winternitz, Classical and Quantum Superintegrability with Applications, {\it J. Phys. A: Math. Theor.},
{\bf 46}, (2013) 423001. (a 97 page topical review paper)


\bibitem{Kress2001} E. G. Kalnins, J. M. Kress,W. Miller Jr. and G. S. Pogosyan,
Completeness of superintegrability in two dimensional constant curvature spaces,
{\it J. Phys. A Math Gen.} {\bf  34}, 4705, (2001).

\bibitem{CK2014} Capel J.J. and  Kress J.M.,
Invariant classification of second-order conformally flat superintegrable systems,
{\it J. Phys.A: Math. Theor}, {\bf 47} (2014), 495202.

\bibitem{Kress2007} J. M. Kress,
Equivalence of superintegrable systems in two dimensions,
{\it Physics of Atomic Nuclei},  {\bf 70}, No. 3, pp. 560–566, (2007).



\bibitem{KMP2007a} E. G.  Kalnins, W. Miller\ Jr.  and S. Post,
         Wilson polynomials and the generic superintegrable system on the 2-sphere,
         {\it J. Phys. A: Math. Theor. \bf 40},  11525-11538 (2007), http://dx.doi.org/10.1088/1751-8113/40/38/005

  \bibitem{LM2014} Q. Li  and W.  Miller Jr., Wilson polynomials/functions and intertwining operators for the generic quantum
superintegrable system on the 2-sphere, 2015 {\it J. Phys.: Conf. Ser.} 597 012059 (http://iopscience.iop.org/1742-6596/597/1/012059)

\bibitem{KMP2014}    E. G. Kalnins, W. Miller\ Jr.  and  S. Post,  Contractions of 2D 2nd order quantum superintegrable systems and the Askey
scheme for hypergeometric orthogonal polynomials
{\em   SIGMA},  {\bf 9} 057, 28 pages, (2013),  http://dx.doi.org/10.3842/SIGMA.2013.057

\bibitem{KM2014}E. G. Kalnins  and  W. Miller Jr.,  Quadratic algebra contractions and 2nd order superintegrable systems,
{\it Anal. Appl.} {\bf 12}, 583-612,  (2014), http://dx.doi.org/10.1142/S0219530514500377


\bibitem{TTW:2001} Tempesta P., Turbiner A.V. and P. Winternitz,
Exact solvability of superintegrable systems,
{\it Journal of Math Physics}, {\bf 42}, (2001), 4248-4257

\bibitem{TurbinerQES2} A.V. Turbiner,
Quasi-exactly-solvable problems and $sl(2)$ algebra,
{\it Comm. Math. Phys.}, {\bf 118}, (1988) 467\,.

\bibitem{Ushveridze} A.G.Ushveridze,
Quasi-exactly solvable models in quantum mechanics
Institute of Physics, Bristol, 1993.

\bibitem{ExactAndQES} E.G. Kalnins, W. Miller, Jr. and and G.S. Pogosyan,
Exact and quasi-exact solvability of second order superintegrable quantum systems. I. Euclidean space preliminaries,
{\it J. Math. Phys.}, {\bf 47}, 033502 (2006); Exact and quasi-exact solvability of second order superintegrable quantum systems. II. Connection with separation of variables, {\it J. Math. Phys.}, {\bf 48}, 023503 (2007)


\bibitem{Turbiner:1988} W. R\"{u}hl and A. V. Turbiner,
Exact solvability of the Calogero and Sutherland models,
{\it Mod. Phys. Lett.}, {\bf A10}, (1995), 2213-2222





\bibitem{Mos} A. Mostafazadeh, Pseudo-Hermiticity versus PT symmetry: the necessary condition for the reality of the spectrum of a non-Hermitian Hamiltonian,
{\it J. Math. Phys.}, {\bf  43}, (2002), 205-214, math-ph/0107001.


\bibitem{SWpotential} J. Fris, V. Mandrosov, Ya. A. Smorodinsky, M. Uhlir and P. Winternitz,
On higher symmetries in quantum mechanics,
{\it Phys.Lett.}, {\bf 16}, (1965), 354-356

\bibitem{Couplingconstant} W. Miller Jr., E. G. Kalnins and S. Post,
Coupling constant metamorphosis and Nth order symmetries in classical and quantum mechanics,
{\it J. Phys. A: Math. Theor.}, {\bf 43} (2010) 035202. (20 pages)



%ccccccccccccccccccccccccccccccccccccccccccccccccccccccccccccccccccccccccccccccccccccccccccccccccccccccccccccccccccccccccc


%
%
%\bibitem{Bromwich} J.T.A. Bromwich, Quadratic forms and their classification by means of invariant factors, Cambridge Tracts \# 3,
%Cambridge University Press, 1904.
%
%\bibitem{Koenigs}
%Koenigs, G., Sur les g\'eod\'esiques a int\'egrales quadratiques. A note
%appearing in ``Lecons sur la th\'eorie g\'en\'erale des
%surfaces''. G. Darboux. Vol 4, 368-404, {\it Chelsea Publishing} 1972.
%
%\bibitem{KM2014}   Kalnins E. G. and Miller, W. Jr., Quadratic algebra contractions and 2nd order superintegrable systems,  {\it Anal. Appl.}
%{\bf 12}, 583-612,  (2014). DOI: 10.1142/S0219530514500377.
%
%\bibitem{Post2011a} Post S., Coupling Constant Metamorphosis, the St\"ackel Transform and Superintegrability, in Symmetries in Nature: Symposium in Memoriam Marcos Moshinsky (Cuernavaca, Mexico, August 9-14, 2010),
%Vol. 1323, Editors L. Benet,P. Hess, J. Torres, K. Wolf, AIP Conference Proceedings, 2011, 265?274.
%
%\bibitem{Post2011b}
% Post S., Models of quadratic algebras generated by superintegrable systems in 2D, SIGMA {\bf 7}
%(2011), 03
%
%\bibitem{CKP2015} J. Capel, J. Kress, S. Post, Invariant Classification and Limits of Maximally Superintegrable Systems in 3D,
%arXiv:1501.06601 [math-ph], 2015
%
%\bibitem{KKMW}
%Kalnins E.\ G., Kress J.\ M., Miller, W.\ Jr. and Winternitz P.,
%{\it Superintegrable systems in Darboux spaces.}
%{\it J.~Math.~Phys.},\  V.44, 5811--5848,  (2003).
%
%\bibitem{Kalnins} E.G. Kalnins,
%Separation of Variables for Riemannian Spaces of Constant
%Curvature,
% Pitman, Monographs and Surveys in Pure and Applied Mathematics
%{\bf 28},
% Longman, Essex, England,
%1986
%
%\bibitem{MPW2013}  Miller, W. Jr.,   Post, S. and  Winternitz, P..  Classical and Quantum Superintegrability with Applications ,
% {\it  J. Phys. A: Math. Theor.} {\bf 46}, (2013) 423001.
%
%\bibitem{Bocher} B\^ocher, M., Ueber die  Reihenentwickelungender Potentialtheorie, B. G. Teubner, Leipzig 1894.
%
%\bibitem{KMR1984}  E.G. Kalnins, W, Miller, Jr,  and G.J. Reid, Separation of variables for complex Riemannian spaces of constant curvature. I.
%Orthogonal separable coordinates for Snc and Enc, {\it Proc. R. Soc. Lond. A }, {\bf 394}, (1984), pp. 183-206.
%
%\bibitem{Miller1977} W. Miller, Jr.,
% Symmetry and Separation of Variables,
% Addison-Wesley,
% Reading, Massachusetts,
%1977
%
%\bibitem{KKM20061}
%E.~G.~Kalnins, J.~M.~Kress and  W.~Miller Jr.
%Second  order superintegrable systems in conformally
%flat spaces.  V: 2D and 3D quantum systems. {\it
%  J. Math. Phys.}, 2006, V.47, 093501.
%
%\bibitem{KMP2010} E. G. Kalnins, W. Miller, Jr  and S. Post,
%Coupling constant metamorphosis and Nth order symmetries in classical and quantum mechanics,
%{\it  J. Phys. A: Math. Theor.}, {\bf 43} (2010) 035202.
%
%\bibitem{4} E.\ G\ Kalnins, J.\ M.\ Kress and W.\ Miller Jr. Second order superintegrable systems in conformally flat spaces I. 2D
%classical structure theory. {\it J.  Math.  Phys.}, {\bf 46}, 053509, (2005).
%
%\bibitem{5} E.\ G.\ Kalnins, J.\ M.\ Kress and W.\ Miller Jr. Second order superintegrable systems in conformally flat spaces
%II. The classical 2D St\"ackel  transform. {\it J.
% Math. Phys,}, {\bf 46}, 053510, (2005).
%
%\bibitem{7} E.\ G.\ Kalnins, J.\ M.\ Kress,W.\ Miller Jr.\  and G.\ S.\ Pogosyan. Completeness of superintegrability in two dimensional constant curvature
%spaces. {\it J. Phys. A Math Gen.} {\bf  34}, 4705, (2001).


\end{thebibliography}
\end{document}